# IMPACT OF THERMAL BEHAVIOR ON OFFSET IN A HIGH-Q GYROSCOPE

*Fei Duan, Jiwei JIAO\*, Yuelin Wang*

The State Key Laboratory of Transducer Technology, Shanghai Institute of Microsystem and Information Technology, Chinese Academy of Sciences

**ABSTRACT**

In this paper, the CFD approach is used to simulate the thermal behavior in a sensitive high-Q gyroscope. The electromagnetically driving wires, in which the alternating current flows, are treated as Joule heat sources in the model. We found that the differences of the temperature, pressure and velocity along the driving direction and transversely across the proof masses increased as the gap height between the proof mass and top glass became smaller. Local pressure gradient is expected to possibly enhance the impact of any imperfect led by MEMS processes or designs on the offset of our tuning fork type gyroscope, which has been experimentally verified. A device with 200um gap gives a two-third offset down compared with that of its counterpart with 50um gap.

**Keywords**- Offset, gyroscope, CFD, pressure gradient

In this paper, the thermal behavior of our electro-magnetically driven high-Q gyroscope is investigated by using the Computational Fluid Dynamics (CFD) approach and we try to understand the related error mechanisms of the offset, a key performance parameter for the microgyroscope. The electromagnetically driving brings large thermal effects, which differs from the electrostatic driving, for the great heat is generated in the so limited space inside the gyroscope. The thermal differences that can be caused by the variation of the pivotal structure parameter are assumed to enhance the impact of the imperfections and further affect the offset, which is experimentally confirmed by the measurement of the zero rate offset.

In section II, our tuning-fork type high-Q microgyroscope and the offset background is briefly described. In section III, we demonstrate the microfabrication process which forms the electromagnetically driving and the capacitive sensing of the gyroscope. In section IV, we then give details about the CFD modeling and show the simulation results of the thermal behavior. The experiment and the measured data are described in section V. Conclusion forms the last section.

## 1. INTRODUCTION

The microelectro-mechanical gyroscopes have attracted great interest in the past decade with a lot of application, such as the Inertial Navigation Systems (INS), platform stabilization, automotive, virtual reality, etc. All these applications are with a need for the stability and accuracy of the microgyroscopes. So the errors of the gyroscopes should be suppressed as few as possible to obtain the perfect output up to the design. The involved cases were studied and a few error mechanisms have been proposed. The cross coupling errors and the parasite Coriolis force were found to be the main error sources [1][2][3]. The interference of the acceleration in the axis direction is also regarded as one of the error sources [4]. It is pointed out that structure imperfection led by microfabrication processes is one of the most important error sources[5][6].

## 2. MICROGYROSCOPE

A novel tuning-fork type microgyroscope with high quality factors in both driving and sensing is presented in this section[7]. The slide-film damping instead of the squeeze-film damping is applied in this system and results in high-Qs, measured about 1000 and 500 in the driving and sensing, respectively. So the microgyroscope can work in the atmosphere and the vacuum packaging is not in demand. Fig.2 shows the schematic and SEM view of the microgyroscope. It can be seen that the gyroscope "symmetrically" consists of two silicon oscillating frames, inside of which proof masses with bar structure detection electrodes are connected by suspension beams, meaningly, it is a tuning-fork type device.

For the tuning-fork type microgyroscope, its offset in the capacitive output, as the capacitive detection is most widely used, is mainly composed of two parts: the electrical part and the mechanical one. The former is produced by the driving and detecting circuits; the latter is usually caused by the structure imperfections led by MEMS processes or designs. For example, the imbalance of the two proofmasses, the overetching in micro-fabrication with uneven bottom of the gyroscope, and so on all lead to the offset. To obtain the reasonable output, the offset is wanted to be stable. However, to compute out the offset of the microgyroscope directly is difficult, for it is caused by so many factors. So we calculate the thermal parameters by simulation to investigate the thermal behavior of the gyroscope first.

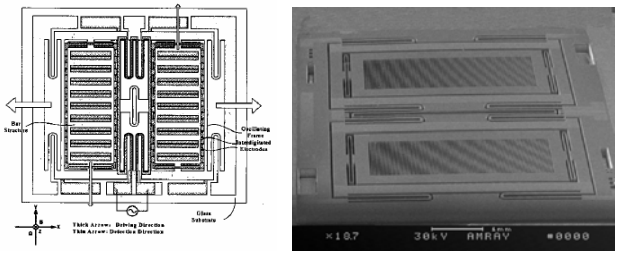

(a) Schematic view  (b)  SEM view
Fig.1 Schematic and SEM view of the microgyroscope

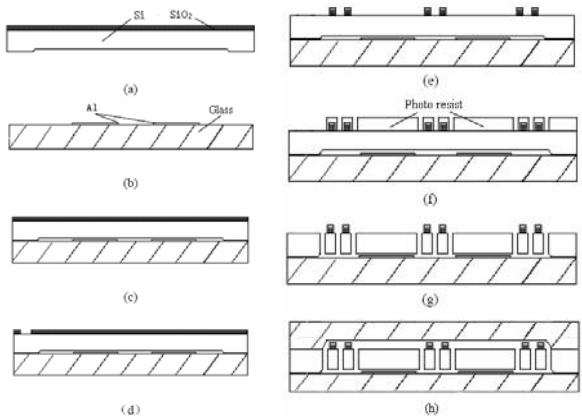

Fig.2  MicrofabricationpProcess flow

(a) Deep cavity etched on a silicon wafer with thermal $SiO_2$ layer; (b) Al interdigitated electrodes patterned on Pyrex7740 glass wafer;  (c) Silicon and glass wafer anodically bonded;  (d) Al driving wire on SiOz layer and ground pad on siliconformed; (e) Oscillating frames and sensing masses patterned;  (f) DRIE to release oscillating frames and sensing masses;  (h) Silicon and deep cavity etched glass wafer bonded.

### 3. MICROFABRICATION

A short overview of the microfabrication process steps is given. The sensor structure is fabricated using silicon bulk micromachining technologies such as ICP deep etching and anodic bonding between silicon and glass wafers. The process flow is shown as Fig.2. The chip is packaged with a permanent magnet fixed above it at atmospheric pressure.

### 4. CFD MODELING

A simplified 2D geometry model with the vertical view is established, as shown in Fig.3. It is a glass-silicon-glass sandwich structure. Aluminum is sputtered on the coupling beam to from the driving wires. When the whole is put in a magnetic field and the alternating current is running through the driving wires, as the function of the Lorentz force, the oscillating frames are vibrating in the plane and thus the gyroscope is driven. The driving wires, as the green shows in Fig.3, are treated as the Joule heat source. The distance between the proofmass and the top galss is defined as the gap height. Simulation is taken with the CFD approach to see the change of the thermal parameters, typically, the temperature, the pressure as well as the velocity.]

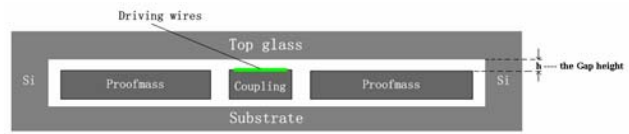

Fig.3 A simplified 2D geometry model of the gyroscope

#### 4.1. Theroy

The thermal parameters are got by numerically solving the energy equation and others. The energy equation is as follows [8]:

$$\frac{\partial}{\partial t}(\rho E) + \nabla \cdot (\vec{\upsilon}(\rho E + P)) = \nabla \cdot \left( k_{eff} \nabla T - \sum_j h_j \vec{J}_j + (\tau_{eff} \cdot \vec{\upsilon}) \right) + S_h$$
(1)

where $k_{eff}$ is the effective conductivity , and $\vec{J}_j$ is the diffusion flux of species j. $S_h$ includes all the defined volumetric heat sources.

In solving the energy equation, for the natural convection flows, we get fast convergence with Boussinesq model than we can get by setting up the problem with fluid density as a function of temperature. The model treats density as a constant value in all solved equations, except for the buoyancy term in the momentum equation:

$$(\rho - \rho_0)g \approx -\rho_0 \beta(T - T_0)g \quad (2)$$

where $\rho_0$ is the constant density of the flow, $T_0$ is the operating temperature, and $\beta$ is the thermal expansion coefficient. $\rho$ is eliminated from the buoyancy term by using the Boussinesq approximation.

**4.2. Simulation results**

The air flow inside the sandwich structure is simulated with a gap between proof masses and top glass. From the model we can get the information of the temperature, pressure, and velocity in the limited space. Fig.4 gives typical simulation results of the velocity vector. In Fig.5, the simulation results of the pressure distributions along the driving direction are presented, corresponding to the different gap heights of 50μm, 100μm, 150μm and 200μm, respectively. The results are concluded in Fig.6 and indicate the dependence of the temperature and the pressure on the gap height. In other words, both the temperature and pressure increases as the gap height goes down. Meanwhile, in Fig.7 we can find that there is a pressure gradient from the center, the heat source, to the side of the device transversely cross the proof mass surfaces. This gradient for large gap is smaller than that for narrow gap. Pressure differences between above and under the proof masses are displayed in Fig.8. The pressure discrepancy of the 50μm gap height is much larger than that of the 150μm gap height, which indicates the greater torque on the proofmass. Furthermore, the pressure difference inside the gyroscope shows similar trend at different environmental temperatures as shown in Fig.9, but the pressure changes more smoothly as the gap is wider.

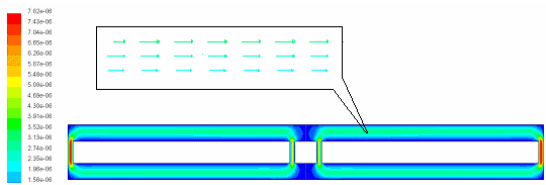

Fig.4 The velocity vectors of the air flow in the gyro

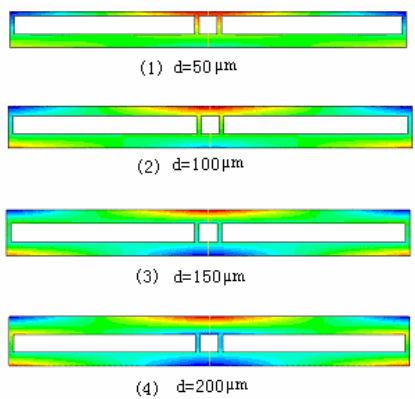

Fig.5 Simulation results of the pressure distribution inside the gyroscope with different gap heights (Red and blue indicate high pressure and low pressure, repectively)

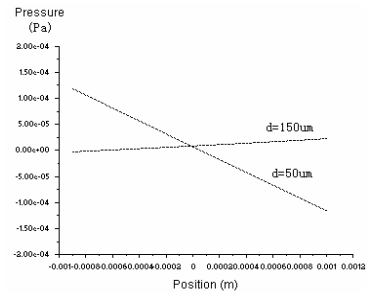

Fig.6 Pressure distribution above proof mass

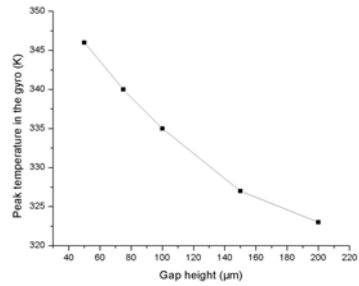

(a) Temerature vs. gap height

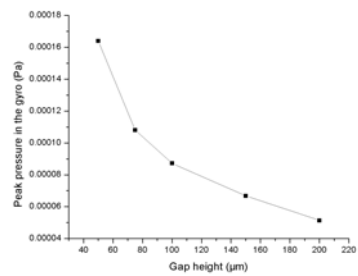

(b) Pressure vs. gap height

Fig.7 Simulation results of temperatures and pressures vs. gap heights

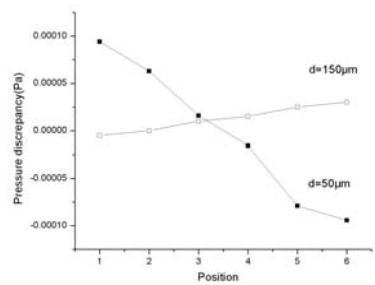

Fig.8 Pressure difference between above and under proofmass

**5 EXPERIMENT**

A series of microgyroscopes with different gap heights are designed and fabricated and their zero point offset are measured. The temperature dependence of the offset for the microgyroscopes with different gap heights is shown in Fig.8. It can be found that the offset dramatically becomes worse in two-third as the gap height goes down from 200um to 50um. The offset of the gyroscope with a 200um gap height is much more stable than that of the gyroscope with a 50um gap height.

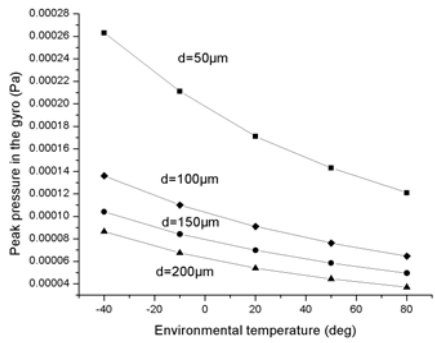

Fig.9 Pressure difference at different environmental temperature

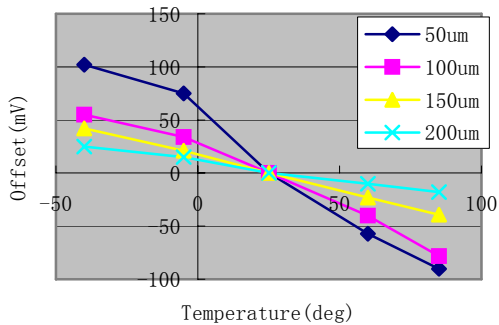

Fig.10 Measured temperature dependence of the offset for different gap heights

The measurement data agree with the simulation results. The same trend of their variation indicates that the temperature dependence of the offset is greatly affected by the thermal difference that can be caused by different gap heights. These differences are expected to possibly enhance the impact of any imperfection led by MEMS processes or designs on the offset. Proper gap height (≥150 microns ) in design is advisable for the electromagnetically driven gyro to obtain the stable offset output. If possible, keep the appropriate environmental temperature.

**6  CONCLUSION**

From the simulation and experimental results, a novel error mechanism for the offset is proposed in our electromagnetically driven high-Q gyroscope. The existence of pressure gradient and its dependence on gap height lead to different torques on the proof masses, which can never be perfect and symmetric, and thus the differential zero-point output will drift accordingly. Such impact can also be expected as an enhancement of other already existed error mechanisms, such as non-orthogonal cross-talk or parasite Coriolis effect. However, further investigation and explanation are still promising and to be expected.